\documentclass[useAMS,usenatbib]{mnras}
\usepackage{natbib}
\usepackage{amsmath}
\usepackage{url}
\usepackage{longtable}
\usepackage{aas_macros}
\usepackage{amssymb}
\usepackage{graphicx}
\usepackage{deluxetable}
\usepackage{pdflscape}
\usepackage{multirow}
\usepackage{color}
\newcommand{\ubv}{\protect\hbox{$U\!BV$}}
\newcommand{\bvri}{\protect\hbox{$BV\!RI$}}
\newcommand{\bvrijh}{\protect\hbox{$BV\!RI\!J\!H$}}

\newcommand{\jh}{\protect\hbox{$J\!H$}}

\newcommand{\jhks}{\protect\hbox{$J\!H\!K_{s}$}}

\newcommand{\about}{$\sim\!\!$~}

\def\lsim{\hbox{\rlap{\raise 0.425ex\hbox{$<$}}\lower 0.65ex\hbox{$\sim$}}}
\def\gsim{\hbox{\rlap{\raise 0.425ex\hbox{$>$}}\lower 0.65ex\hbox{$\sim$}}}

\def\arcsec{\hbox{$^{\prime\prime}$}}

\title[Luminosity Differences for Twin SNe~Ia]{Significant Luminosity Differences of Two Twin Type Ia Supernovae}

\def\ucsc{1}
\def\stsci{2}
\def\tamu{3}
\def\jhu{4}
\def\ucb{5}
\def\miller{6}
\def\uw{7}
\def\az{8}

\begin{document}

\author[Foley, et~al.]{Ryan~J.~Foley$^{\ucsc}$\thanks{E-mail:foley@ucsc.edu},
Samantha~L.~Hoffmann$^{\stsci}$,
Lucas~M.~Macri$^{\tamu}$,
Adam~G.~Riess$^{\jhu,\stsci}$,
\newauthor
Peter~J.~Brown$^{\tamu}$,
Alexei~V.~Filippenko$^{\ucb,\miller}$,
Melissa~L.~Graham$^{\uw}$,
Peter~A.~Milne$^{\az}$\\
$^{\ucsc}$Department of Astronomy and Astrophysics, University of California, Santa Cruz, CA 95064, USA\\
$^{\stsci}$Space Telescope Science Institute, 3700 San Martin Drive, Baltimore, MD 21218, USA\\
$^{\tamu}$George P.\ and Cynthia Woods Mitchell Institute for Fundamental Physics and Astronomy, Department of Physics \& Astronomy,\\ Texas A\&M University, College Station, TX 77843, USA\\
$^{\jhu}$Department of Physics and Astronomy, Johns Hopkins University, Baltimore, MD, USA\\
$^{\ucb}$Department of Astronomy, University of California, Berkeley, CA 94720-3411, USA\\
$^{\miller}$Miller Senior Fellow, Miller Institute for Basic Research in Science, University of California, Berkeley, CA 94720, USA\\
$^{\uw}$Department of Astronomy, University of Washington, Box 351580, U.W., Seattle, WA 98195-1580, USA\\
$^{\az}$Steward Observatory, University of Arizona, 933 North Cherry Avenue, Tucson, AZ 85721, USA
}

\date{Accepted  . Received   ; in original form  }
\pagerange{\pageref{firstpage}--\pageref{lastpage}} \pubyear{2016}
\maketitle
\label{firstpage}

\begin{abstract}
  The Type Ia supernovae (SNe~Ia) 2011by, hosted in NGC~3972, and
  2011fe, hosted in M101, are optical ``twins,'' having almost
  identical optical light-curve shapes, colours, and
  near-maximum-brightness spectra.  However, SN~2011fe had
  significantly more ultraviolet (UV; $1600 < \lambda < 2500$~\AA)
  flux than SN~2011by before and at peak luminosity.  Theory suggests
  that SNe~Ia with higher progenitor metallicity should (1) have
  additional UV opacity near peak and thus lower UV flux; (2) have an
  essentially unchanged optical spectral-energy distribution; (3) have
  a similar optical light-curve shape; and (4) because of the excess
  neutrons, produce more stable Fe-group elements at the expense of
  radioactive $^{56}$Ni and thus have a lower peak luminosity.
  \citet{Foley13:met} suggested that the difference in UV flux between
  SNe~2011by and 2011fe was the result of their progenitors having
  significantly different metallicities.  The SNe also had a large,
  but insignificant, difference between their peak absolute magnitudes
  ($\Delta M_{V, {\rm ~peak}} = 0.60 \pm 0.36$~mag), with SN~2011fe
  being more luminous.  We present a new Cepheid-based distance to
  NGC~3972, significantly improving the precision of the distance
  measurement for SN~2011by.  With these new data, we determine that
  the SNe have significantly different peak luminosities ($\Delta
  M_{V, {\rm ~peak}} = 0.335 \pm 0.069$~mag), corresponding to
  SN~2011fe having produced 38\% more $^{56}$Ni than SN~2011by, and
  providing additional evidence for progenitor metallicity differences
  for these SNe.  We discuss how progenitor metallicity differences
  can contribute to the intrinsic scatter for
  light-curve-shape-corrected SN luminosities, the use of ``twin'' SNe
  for measuring distances, and implications for using SNe~Ia for
  constraining cosmological parameters.
\end{abstract}



\defcitealias{Foley13:met}{FK13}
\defcitealias{Graham15}{G15}
\defcitealias{Phillips99}{P99}

\section{Introduction}\label{s:intro}

Type Ia supernovae (SNe~Ia) are excellent standardisable candles that
can be measured to cosmological distances.  Observations of SNe~Ia
have been crucial in discovering cosmic acceleration
\citep{Riess98:Lambda, Perlmutter99}, as well as making the most
precise measurements of the Hubble constant, $H_{0}$
\citep[e.g.,][]{Riess16}, and the equation-of-state parameter for dark
energy, $w$, \citep[e.g.,][]{Jones18, Scolnic18:ps1}.

SNe~Ia are {\it not} ``standard candles'' (just as Cepheid variables
are not standard candles, strictly speaking), having a factor of
\about10 difference in luminosity from one extreme to the other.
However, after correcting for light-curve shape
\citep[e.g.,][]{Phillips93} and colour \citep[e.g.,][]{Riess96,
  Tripp98}, SNe~Ia have a small scatter in their measured distances
\citep[typically \about 8\%; e.g.,][]{Hicken09:de, Stritzinger11}.  A
corollary to this empirical measurement is that two SNe~Ia with
exactly the same light-curve shapes, colours, and spectra should have
luminosities that differ by at most the intrinsic luminosity scatter
that belies additional, unaccounted physical diversity.

One possible driver for the nonzero intrinsic scatter is progenitor
stars with differing metallicity.  In this case, the number of
neutrons increases (alternatively, $Y_{e}$ decreases) with increasing
metallicity.  When the star explodes, the additional neutrons result
in more stable Fe-group elements at the expense of radioactive
$^{56}$Ni, which powers the SN light curve and directly affects its
peak luminosity \citep{Timmes03, Bravo10}.  However, since roughly the
same amounts of Fe-group elements are generated, the overall optical
opacity is roughly the same, and the light-curve shape and colour are
unaffected \citep{Mazzali06}.  Consequently, differing progenitor
metallicity could cause the residual luminosity scatter observed among
SNe~Ia.

In addition to causing larger statistical distance uncertainties,
changing progenitor metallicity with redshift could systematically
bias the measurement of cosmological parameters
\citep{Podsiadlowski06}.  Specifically, if the average progenitor
metallicity decreases with increasing redshift, one might expect a
drift in the mean peak luminosity of SNe~Ia with redshift.
Determining if such an effect exists, its magnitude, and its
relationship with metallicity is critical for precisely and accurately
measuring cosmological parameters with SNe~Ia.

While changing progenitor metallicity has a minimal effect on the
optical light curves, colours, and spectra of SNe~Ia, the ultraviolet
(UV) is significantly affected \citep[e.g.,][]{Hoflich98, Lentz00,
  Sauer08, Walker12}.  In most models, increasing the progenitor
metallicity increases the line blanketing within the outer layers of
the ejecta.  While these layers quickly become optically thin at
longer wavelengths, they remain optically thick to UV photons through
peak brightness.  As a result, higher metallicity progenitors result
in SNe~Ia with depressed UV flux relative to the optical flux.

Differences in the UV continuum for low- and high-redshift SNe~Ia have
been detected in different surveys \citep{Foley12:sdss, Maguire12,
  Milne15}.  While these detections are consistent with changing
progenitor metallicity with redshift, there could be other effects
related to changing populations that do not bias cosmological
measurements.

The current sample of low-redshift SNe~Ia with high-quality UV
spectroscopy is relatively small \citep{Kirshner93, Foley12:11iv,
  Foley14:14j, Foley16:uv, Foley13:ca, Foley13:met, Mazzali14,
  Pan15:13dy}.  A much larger sample of UV spectra from {\it Swift}
has recently been published \citep{Pan18}, but the spectra are of
lower quality than what is typically obtained with the {\it Hubble
  Space Telescope} ({\it HST}).  None the less, the sample of SNe~Ia
with UV spectra has been critical for understanding the connection
between progenitor metallicity, UV properties, and optical luminosity.

In particular, the best indication of different progenitor metallicity
for two SNe~Ia comes from our studies of SNe~2011by and 2011fe
(\citealt{Foley13:met}; \citealt{Graham15}; hereafter
\citetalias{Foley13:met} and \citetalias{Graham15}, respectively;
although see also \citealt{Milne13}, \citealt{Mazzali14}, and
\citealt{Foley16:uv}), two ``twin'' SNe~Ia \citep[e.g.,][]{Fakhouri15}
with nearly identical optical light curves, colours, and spectra, but
dramatically different UV colours and spectra.  Because of their
similar optical properties, the differences in UV behaviour can be
constrained to progenitor metallicity \citepalias{Foley13:met}, with
the SN~2011by and SN~2011fe progenitors being above and below solar
metallicity, respectively.  While the SNe have different late-time ($t
> 100$~days) optical decline rates, their nebular spectra are nearly
identical \citepalias{Graham15}.

Having established that SNe~2011by and 2011fe likely had substantially
different progenitor metallicity, they provide an excellent
opportunity to directly determine the effect of metallicity on SN
luminosity and distance estimates.  Originally,
\citetalias{Foley13:met} determined that their peak $V$-band absolute
magnitudes differed by \about 0.6~mag, significantly larger than the
intrinsic scatter of SNe~Ia.  However, this difference directly
depends on the distances assumed for these nearby SNe.  In particular,
while there was a measured Cepheid distance to M101, the host galaxy
for SN~2011fe \citep{Shappee11}, there previously was only a
Tully-Fisher distance to NGC~3972 \citep{Tully09}, the host galaxy of
SN~2011by.  The latter had a relatively large uncertainty (0.36~mag),
and systematic differences between the Cepheid and Tully-Fisher scale
could have been the main cause for the apparently large differences in
SN luminosity.

To reduce this source of uncertainty, we obtained a series of {\it
  HST} images to find Cepheid variable stars in NGC~3972 and measure
both a precise absolute distance to NGC~3972 and a relative distance
between M101 and NGC~3972, removing uncertainties common to both
measurements.  From this analysis, we find that the distance to
NGC~3972 is similar to the Tully-Fisher measurement (although with a
shift of 0.26~mag), and SNe~2011by and 2011fe do indeed have
significantly different luminosities.

We present our {\it HST} observations in Section~\ref{s:obs}.  We
detail our Cepheid distance estimates, direct luminosity comparisons
of the SNe, and direct SN distance estimates in Section~\ref{s:anal}.
We discuss the implications of this result and conclude in
Section~\ref{s:disc}.


\section{Observations}\label{s:obs}

\subsection{Cepheids in NGC~3972}

To obtain a precise Cepheid distance to NGC~3972, we observed the
galaxy using the {\it HST} Wide-Field Camera 3 (WFC3) UVIS and
infrared (IR) channels (Programme GO--13647; PI Foley).  Details of
the UVIS and IR observations and data reduction are presented by
\citet{Hoffmann16} and \citet{Riess16}, respectively.  Here we briefly
review them.

Observations with WFC3/UVIS and the F350LP ``white light'' filter were
used to discover Cepheids and measure their periods.  The data were
obtained over 12 separate epochs, with the specific timing chosen to
minimise the integral of the power over the frequency interval
corresponding to the observation window.  For 6 epochs, we also
obtained observations with WFC3/IR and the F160W (roughly $H$) filter.
For the other 6 epochs, half had an observation with WFC3/UVIS and the
F555W (roughly $V$) filter and the other half had an observation with
WFC3/UVIS and the F814W (``wide $I$'') filter.  The filtered
observations are used primarily to aid in selection and to constrain
dust reddening.

\begin{figure*}
\begin{center}
\includegraphics[angle=0,width=6.4in]{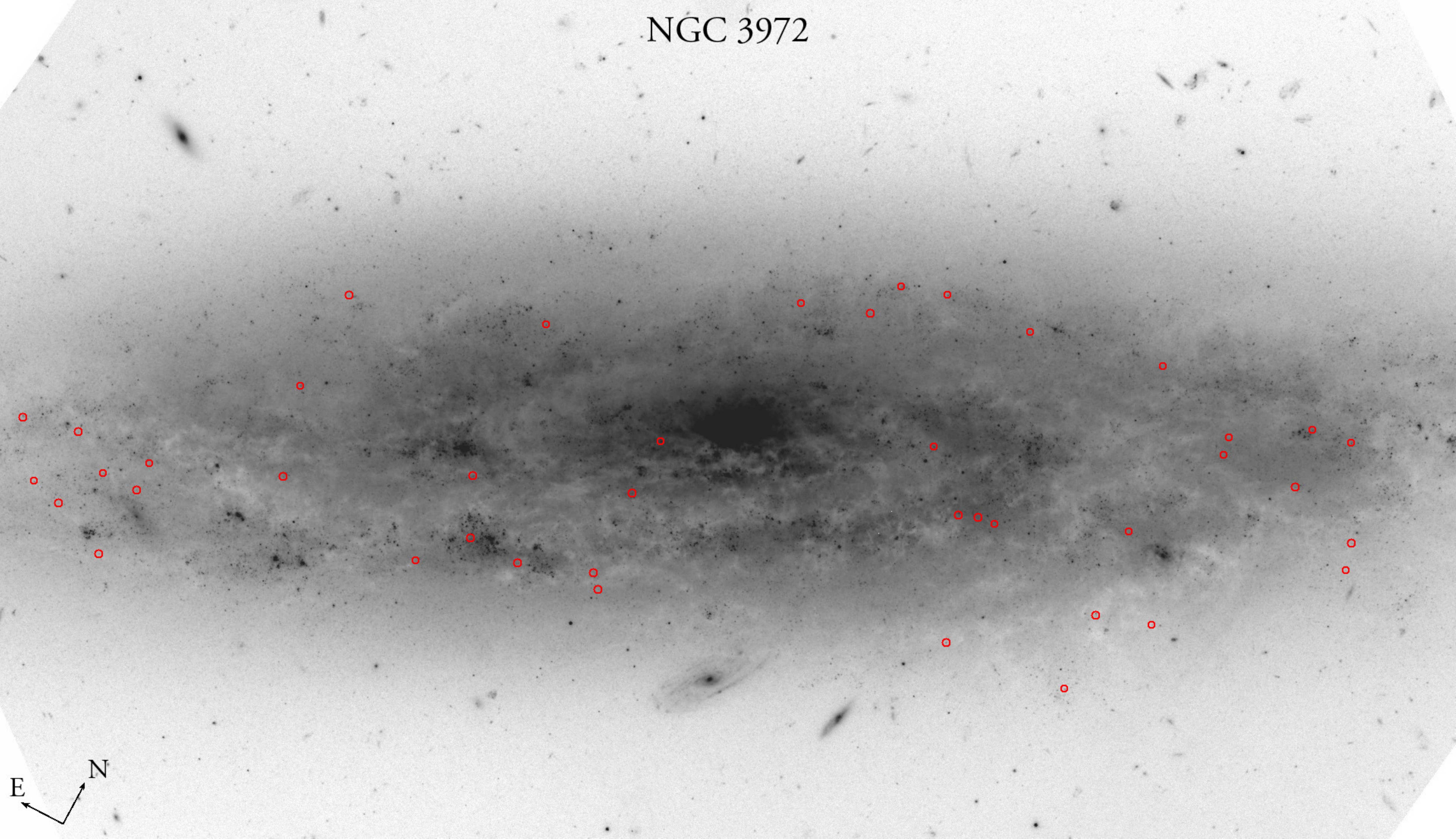}
\caption{90\arcsec\ $\times$ 160\arcsec\ {\it HST}/WFC3 F350LP image
  of NGC~3972.  The image is generated from the stack of all F350LP
  data and the intensity scale is logarithmic.  To aid orientation, a
  compass rose is shown, with each arrow corresponding to 5\arcsec.
  The positions of Cepheid variables used to determine the distance to
  NGC~3972 are marked with red circles.}\label{f:finder}
\end{center}
\end{figure*}

The first and last epochs were obtained on 19 April 2015 and 8 July
2015 UT, respectively, corresponding to an 80-day baseline.  Data were
retrieved from the Mikulski Archive for Space Telescopes (MAST) and
reduced as described by \citet{Hoffmann16} and \citet{Riess16}.  An
{\it HST}/WFC3 image of NGC~3972 is shown in Figure~\ref{f:finder}.

Photometry was performed with {\tt DAOPhot}/{\tt ALLSTAR}
\citep{Stetson87} and {\tt ALLFRAME} \citep{Stetson94} using
point-spread functions (PSFs) created with {\tt TinyTim}
\citep{Krist11} as described by \citet{Macri06},
\citet{Riess09:cepheids}, \citet{Riess11}, and \citet{Hoffmann16}.

\subsection{{\it HST} Spectra of SN~2011by}

Optical spectra of SN~2011by were obtained with {\it HST}/STIS
(Programme GO--12298; PI Ellis).  STIS produces excellent absolute and
relative spectrophotometry, and therefore the relative luminosity as a
function of wavelength for similar-phase spectra can be used to
determine the relative reddening of two twin SNe.

\begin{figure}
\begin{center}
\includegraphics[angle=0,width=3.2in]{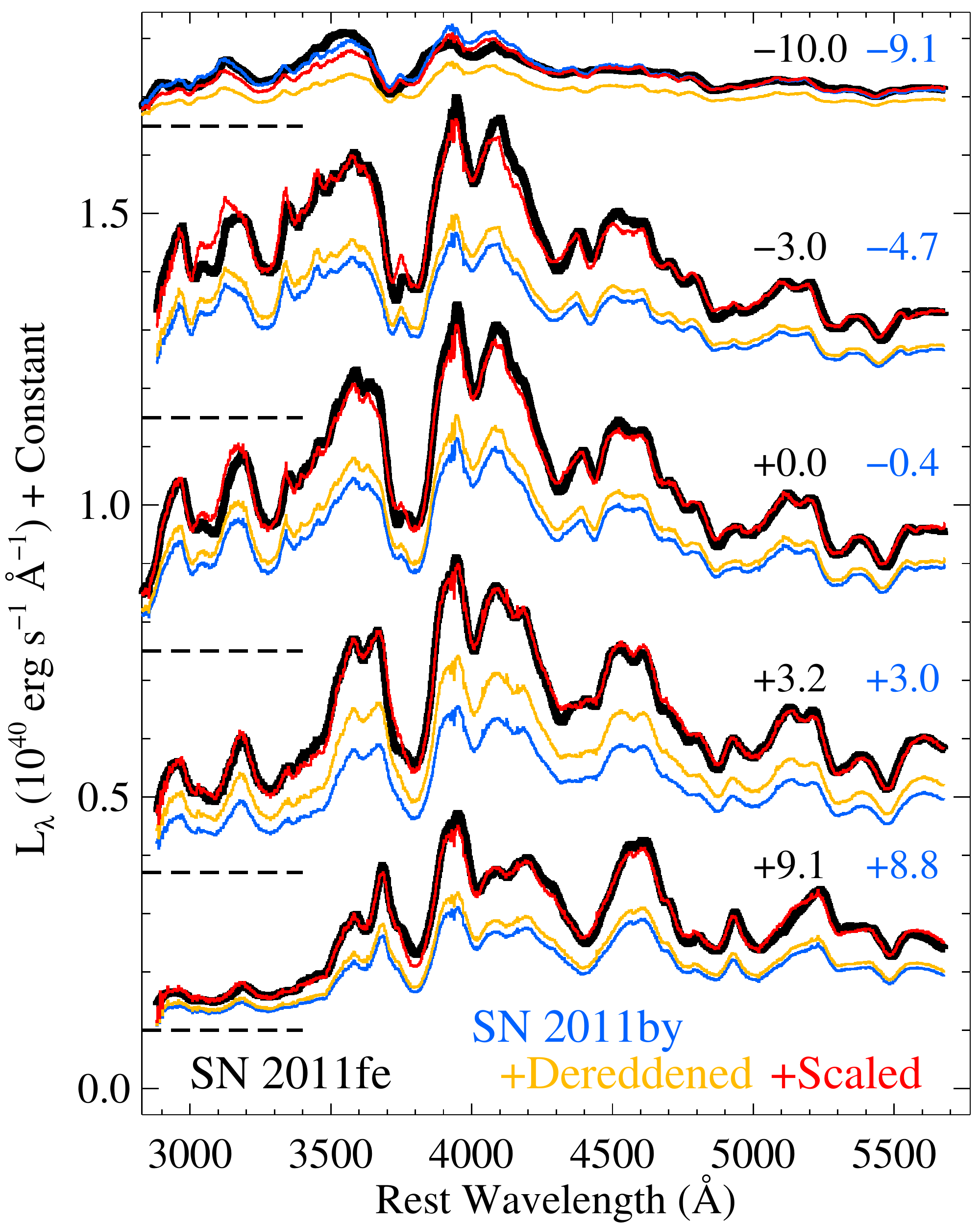}
\caption{{\it HST}/STIS spectra of SNe~2011by (blue curves) and 2011fe
  (black curves) for 5 matched phases with each phase labeled next to
  the spectra.  The spectra have been scaled by their geometric
  dilution factor so that they are in $L_{\lambda}$ units.  The
  spectra have been shifted vertically by arbitrary amounts indicated
  by the dashed lines.  Also plotted are the SN~2011by spectra after
  being dereddened (gold curves) and scaled (red curves) to match the
  corresponding SN~2011fe spectra using the best-fitting values for
  both $E(B-V)$ and the scale factor, as discussed in the text.  For
  all but the first epoch, the similarity in spectral shape and
  features is striking.}\label{f:spec}
\end{center}
\end{figure}

SN~2011by was observed with STIS on 5 epochs, although only two use a
UV setting.  The two UV/optical spectra were presented by \citet[but
only the optical portion]{Maguire12}, \citetalias{Foley13:met}, and
\citetalias{Graham15}.  The remaining spectra are first presented
here.  The data were reduced using the standard \textit{HST} Space
Telescope Science Data Analysis System (STSDAS) routines to
bias-subtract, flat-field, extract, wavelength-calibrate, and
flux-calibrate each SN spectrum.  Similar reductions were performed
for the SN~2011fe spectra used in this study \citep{Foley12:11iv,
  Foley14:14j, Foley16:uv, Foley13:met, Foley13:ca, Pan15:13dy}.  A
log of all observations is presented in Table~\ref{t:11by}.  The
SN~2011by spectra are shown in Figure~\ref{f:spec}.

\begin{deluxetable}{rlr}
\tabletypesize{\footnotesize}
\tablewidth{0pt}
\tablecaption{{\it HST}/STIS Spectral Observations of SN~2011by\label{t:11by}}
\tablehead{
\colhead{Phase\tablenotemark{a}} &
\colhead{UT Date} &
\colhead{Exposure (s)\tablenotemark{b}}}

\startdata

$-9.1$ & 2011 Apr.\ 30.522 & 8300+2263\tablenotemark{c} \\
$-4.7$ & 2011 May    5.139 &    0+2263 \\
$-0.4$ & 2011 May    9.343 & 5316+2263\tablenotemark{d} \\
  3.0  & 2011 May   13.749 &    0+2263 \\
  8.8  & 2011 May   18.203 &    0+2263 \\

\enddata

\tablenotetext{a}{Days since $B$ maximum, 2014 Feb.\ 2.0 (JD
  2,456,690.5).}

\tablenotetext{b}{First and second numbers correspond to the time for
  the G230L and G430L gratings, respectively.}

\tablenotetext{c}{Originally published by \citetalias{Graham15}.}

\tablenotetext{d}{G430L data originally published by
  \citet{Maguire12}; G230L data originally published by
  \citetalias{Foley13:met}.}

\end{deluxetable}

\section{Analysis}\label{s:anal}

\subsection{Selecting Cepheids in NGC~3972}

First, the F350LP light curves of all sources were examined to
determine the subset that are variable.  We visually inspected the
light curves and rejected obviously spurious photometry and photometry
with unusually large uncertainties.  The Welch-Stetson variability
index \citep{Stetson96} was used to determine which objects are
variable.

We matched the light curves of all sources with template Cepheid light
curves \citep{Yoachim09} of periods between 10 and 100~days. All data
(regardless of band) were used for the fitting.  From the best-fitting
period, the model predicts a corresponding amplitude.  Comparing the
predicted amplitude to the F350LP light curve, we measured the
$\chi^{2}$ statistic.  More than 90\% of the objects selected as
variable were poorly matched by the Cepheid template light curves.
For the remaining subset, we further refined the best-fitting
parameters.  Finally, we used additional criteria to remove objects
that are inconsistent with isolated low-to-moderate-reddening
Cepheids.

In total, we selected 71 Cepheids.  We present a finding chart of
NGC~3972 with the Cepheids marked in Figure~\ref{f:finder}.  Details
of the selection process can be found in \citet{Hoffmann16}.

\subsection{Cepheid Distance for NGC~3972}

\begin{figure}
\begin{center}
\includegraphics[angle=0,width=3.2in]{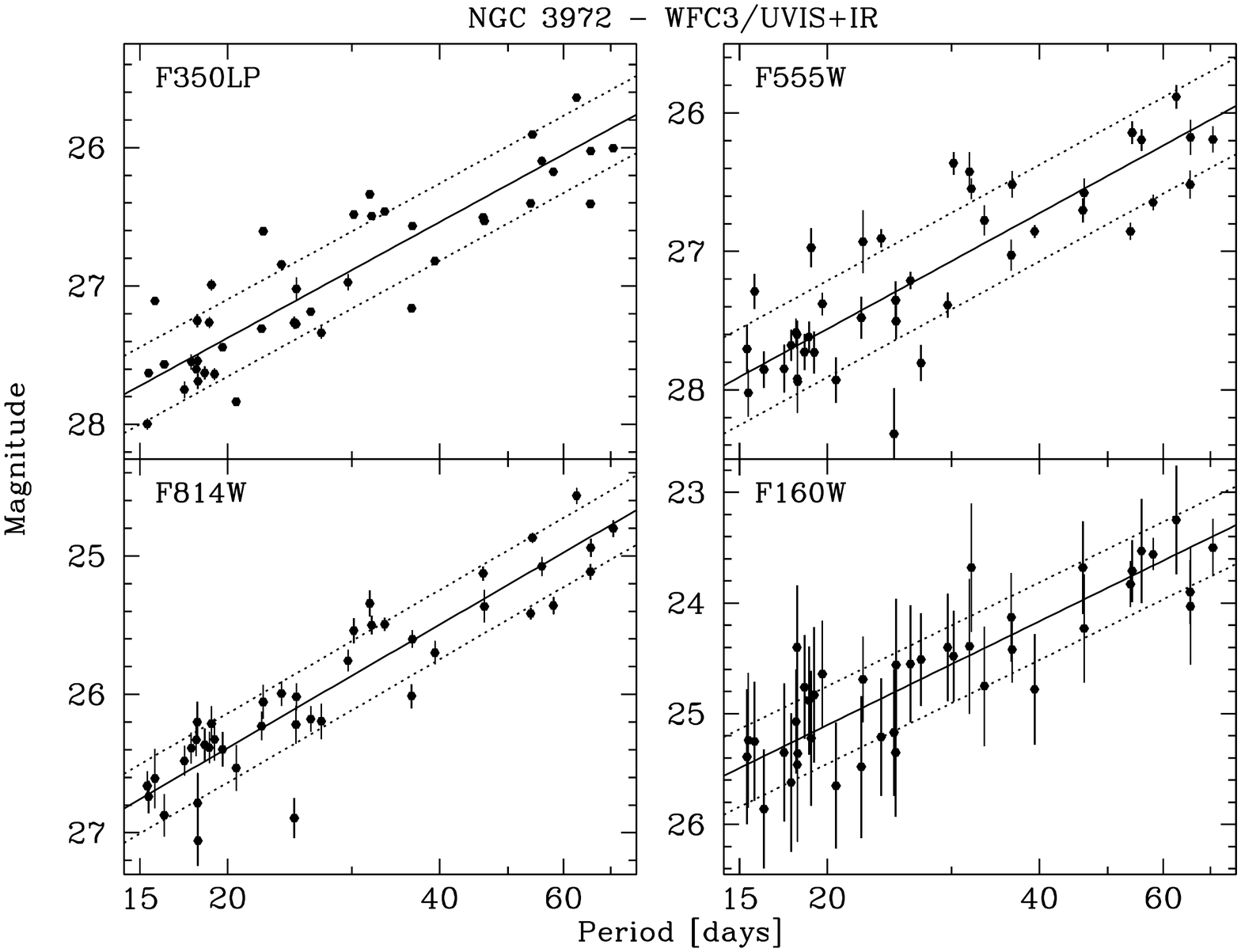}
\caption{Period-luminosity relations (Leavitt laws) for the Cepheid
  variables in NGC~3972 in the {\it HST}/WFC3 F350LP (upper left),
  F555W (upper right), F814W (lower left), and F160W (lower right)
  filters.}\label{f:pl}
\end{center}
\end{figure}

As was done by \citet{Riess16}, we use the period-luminosity relation
(Figure~\ref{f:pl}), correcting for metallicity, and a combination of
NGC~4258, the Milky Way, and the Large Magellanic Cloud (LMC) as an
anchor to determine the distance of NGC~3972.  We refer the reader to
\citet{Riess16} for details of this calculation.

For all measured Cepheids, we determine the Wesenheit magnitude
\citep{Madore82}, which corrects for extinction and the finite
temperature width of the instability strip.  We compute a
period-luminosity relation and remove significant outliers as well as
Cepheids with periods below the completeness limit
\citep[see][]{Riess16}.  Figure~\ref{f:pl} displays the final
period-luminosity relation for NGC~3972, consisting of 42 Cepheids.
The Cepheid data imply a distance modulus of $31.594 \pm 0.071$~mag.

Additionally, we can directly compare the Cepheid data for NGC~3972
and M101 to determine their relative distances without the
uncertainties of the absolute distance scale.  The relative distance
is particularly important for comparing SNe~2011by and 2011fe.
Directly comparing the Cepheid data for both galaxies and accounting
for covariances in the measurements, we determine that NGC~3972 has a
distance modulus that is $2.459 \pm 0.062$~mag greater than that of
M101\footnote{We use the \citet{Riess16} determination of the Cepheid
  distance for M101 ($\mu = 29.14$~mag) rather than that of
  \citet[$\mu = 29.04$~mag]{Shappee11}.  Primarily, this is so that
  both galaxies have consistent methodology.  However, we note that
  the measurements are consistent (to within 0.01~mag) when the same
  LMC or NGC~4258 zeropoints are used.}.

\subsection{Luminosity Differences for SNe~2011by and 2011fe}

With the absolute and relative distances to M101 and NGC~3972, we can
directly compare the observations of SNe~2011by and 2011fe on an
absolute scale.  However, to do this, one must correct for any
host-galaxy extinction.  While both SNe are consistent with having
zero reddening, SN~2011by is slightly redder than SN~2011fe.  In order
to account for the possibility of a different reddening for each SN
and a coherent, grey offset in their absolute magnitudes, we
simultaneously fit for both.

Conveniently, SNe~2011by and 2011fe were both observed by {\it Swift}
\citep{Brown12, Milne13} and the Lick Observatory 0.76~m Katzman
Automatic Imaging Telescope \citep[KAIT;][]{Silverman13,
  Zhang16:11fe}, making $S$-corrections \citep{Stritzinger02}
unnecessary (and thus removing one potential systematic uncertainty).
Unfortunately, near peak brightness, SN~2011fe was too bright to have
accurate flux measurements in the {\it Swift} \ubv\! bands.  As a
result, there is a small phase range of only a few days, far from
peak, where both SNe have {\it Swift} $B$ photometry; it is therefore
difficult to determine a precise offset in this band and these data
are not used in our analysis.  Because of potential intrinsic UV
differences among the two SNe, we also do not use the $U$-band (and
bluer bands) {\it Swift} data to measure the reddening.  However, both
SNe have sufficient overlap in {\it Swift} $V$ to determine a
magnitude offset in that band.

Additionally, \citet{Matheson12} observed SN~2011fe in \jh\ and
\citet{Friedman15} observed SN~2011by in \jhks.  While each SN was
observed with different telescopes, instruments, and slightly
different filters, we were able to determine the necessary
$S$-corrections using the SN~2011fe near-infrared (NIR) spectral
sequence \citep{Hsiao13} and the available filter
functions\footnote{https://www.noao.edu/kpno/manuals/whirc/filters.html}
\citep{Cohen03}.  This process is similar to what was done by
\citet{Weyant18} for comparing SNe~Ia in these two systems.

Using their peak-brightness spectra, we measured a $K$-correction for
each SN.  The relative $K$-correction was $<$0.012~mag for all bands.

\begin{figure}
\begin{center}
\includegraphics[angle=0,width=3.2in]{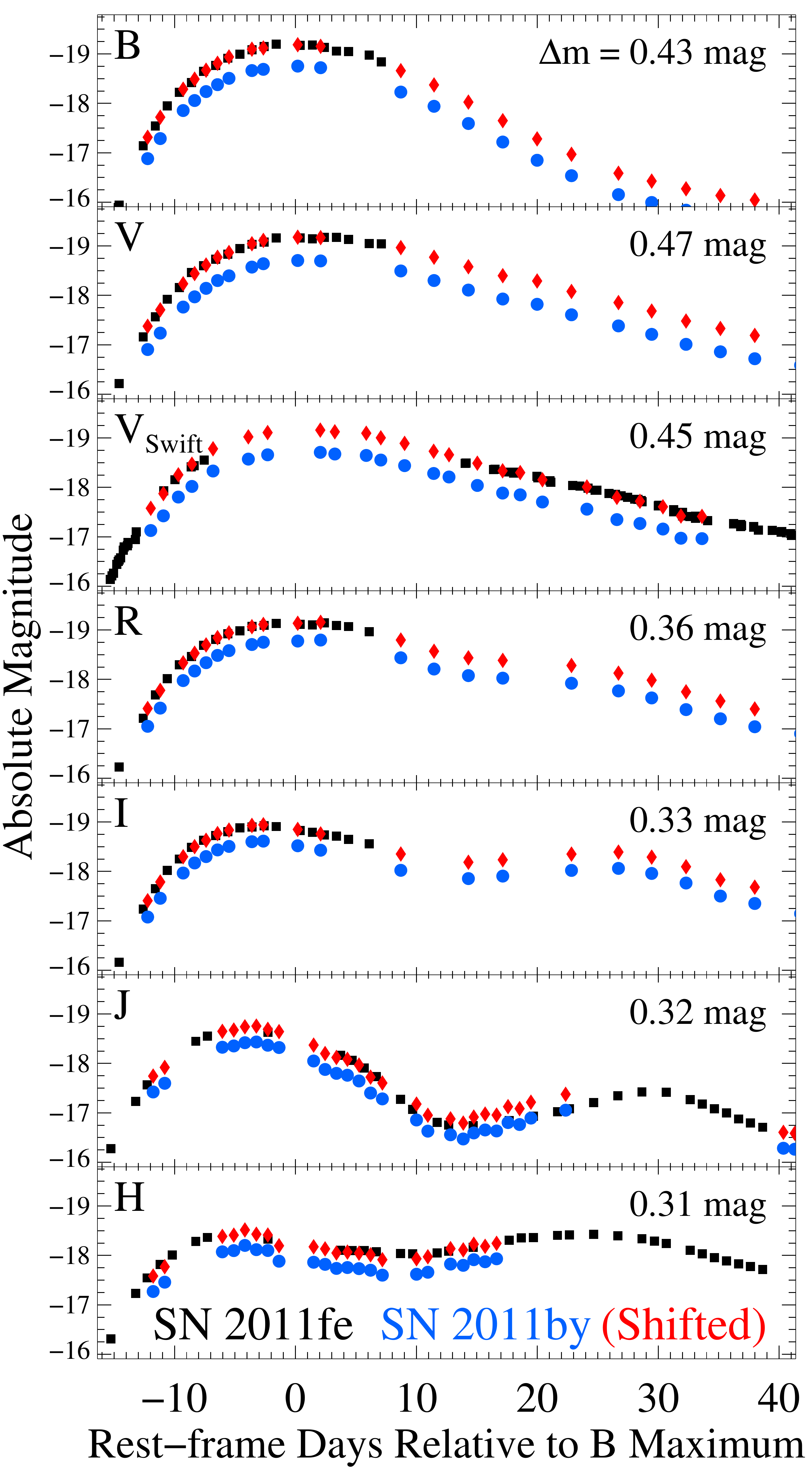}
\caption{Optical and NIR light curves of SNe~2011by (blue circles) and
  2011fe (black squares) in KAIT \bvri, {\it Swift} $V$, and $JH$
  bands.  The red diamonds represent the SN~2011by light curve shifted
  by a single magnitude for each band, noted in each subpanel, to
  match the luminosity of SN~2011fe.}\label{f:lc_shifts}
\end{center}
\end{figure}

\begin{figure}
\begin{center}
\includegraphics[angle=0,width=3.2in]{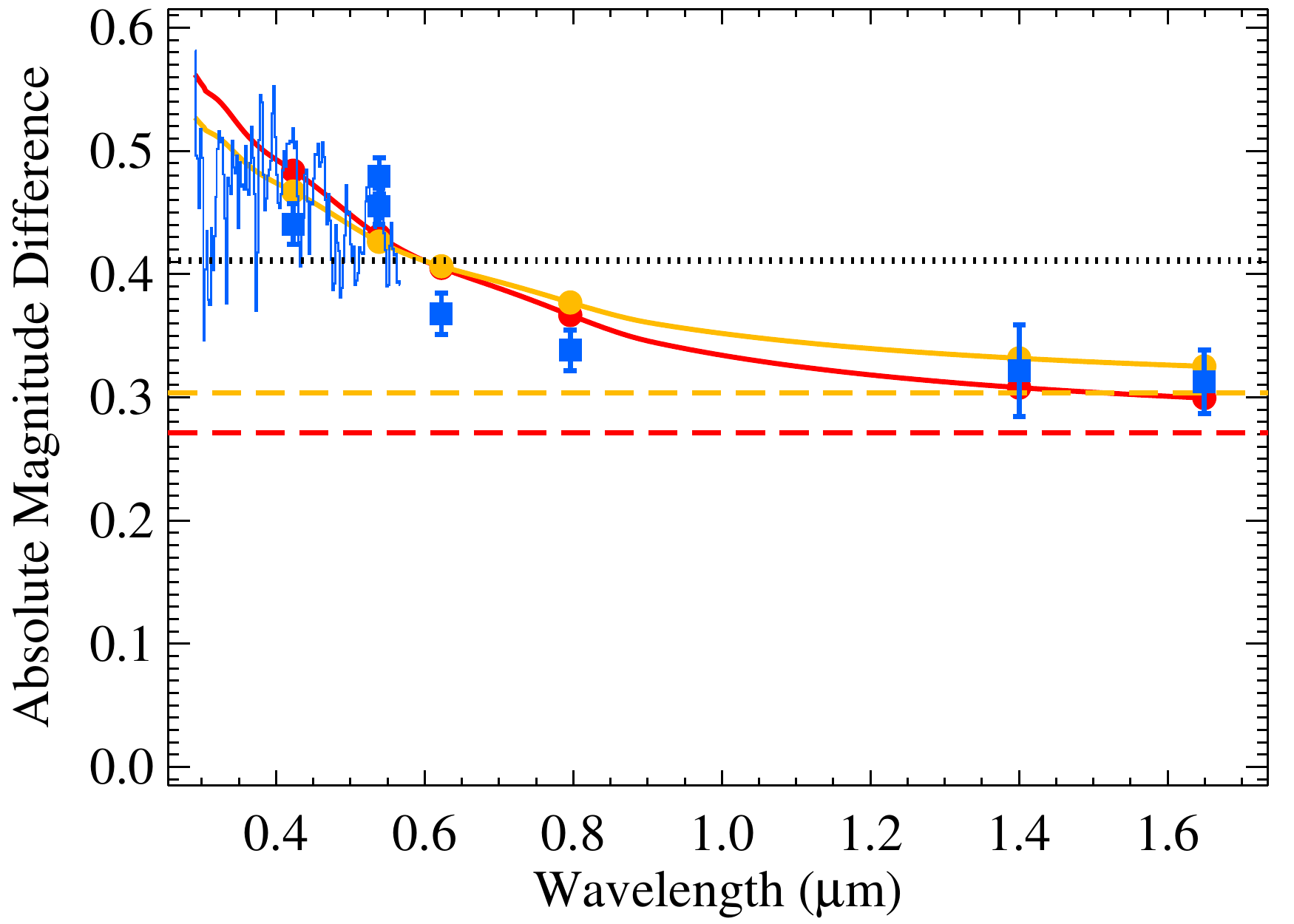}
\caption{Absolute magnitude difference between SNe~2011by and 2011fe
  as a function of wavelength.  The blue squares show the KAIT \bvri,
  {\it Swift} $V$, and $JH$ bands.  The effective wavelength for each
  point is determined by convolving the filter transmission functions
  with the peak-brightness spectrum.  The blue curve represents the
  difference in their peak-brightness {\it HST}/STIS spectra.  The
  black dotted line illustrates the best-fitting constant offset to
  all data.  The red and gold solid lines represent the best-fitting
  reddening and offset model (using a \citealt{Fitzpatrick99}
  reddening law with $R_{V} = 3.1$) when only fitting the photometry
  and varying the reddening, and using spectral matches to determine
  the reddening and the photometry to determine the offset,
  respectively.  The red and gold dashed lines display the
  corresponding constant offsets.  Our best-estimate reddening is
  $E(B-V) = 0.039 \pm 0.006$~mag and best-estimate constant offset is
  $0.304 \pm 0.062$~mag, which corresponds to the gold dashed line.
}\label{f:offset}
\end{center}
\end{figure}

Correcting for Milky Way extinction \citep{Schlafly11},
$K$-corrections, $S$-corrections (for \jh), and their distances, we
produce absolute magnitude light curves in \bvrijh, which we present
in Figure~\ref{f:lc_shifts}.  We note that the absolute magnitude
light curves are not corrected for any potential host-galaxy
extinction.  To determine the relative magnitude offset in each band,
we use a b-spline function to interpolate each light curve to match
the phases for the other light curve.  We then fit for an offset
between the two light curves.  These offsets, as a function of
effective filter wavelength (as determined from the peak-brightness
SN~2011fe spectrum), are presented in Figure~\ref{f:offset}.  In all
bands, SN~2011fe is 0.31 to 0.48~mag more luminous than SN~2011by,
with bluer bands having a larger offset.  As a cross-check,
\citet{Dhawan18} found a difference in the peak absolute $J$ magnitude
of 0.30~mag, comparable to our full light-curve $J$-band offset of
0.32~mag, using the same data.  The average offset in all bands is
$0.411 \pm 0.007$~mag, but a single offset for all bands does not fit
the data well, having $\chi^{2}/{\rm dof} = 80.1/6$.

Notably, the offset between SNe~2011by and 2011fe in a given band is
similar at all epochs probed by the photometry.  The one exception is
for the $J$ band for phases between \about 10 and \about 25~days after
peak brightness.  This indicates that the intrinsic colour evolution
for the SNe is similar and that the offsets in a given band are
primarily caused by slow-changing attributes such as a luminosity
offset and dust extinction.

Assuming that the change in absolute magnitude offset with wavelength
is caused by differential dust reddening, we simultaneously fit for a
constant offset and dust reddening.  Using either a \citet{Cardelli89}
or \citet{Fitzpatrick99} reddening law, and fixing $R_{V}$ to be 3.1,
we find consistent reddening values with $E(B-V) \approx 0.047$~mag.
If we allow $R_{V}$ to float, then we find best-fit values for
$E(B-V)$ of \about 0.038~mag with $R_{V} \approx 4.2$.  Regardless of
the choice of reddening law, the constant offset is \about 0.27~mag.
For our final analysis, we use the \citet{Fitzpatrick99} reddening
law.

To further constrain the reddening, we examine the spectrophotometry
of SNe~2011by and 2011fe.  Optical spectra of both SNe were obtained
with {\it HST}/STIS.  SN~2011fe was observed on 10 separate epochs
covering phases of $-13$ to $+40$~days \citep{Maguire12, Foley13:ca,
  Foley13:met, Mazzali14}.

We match each SN~2011by spectrum to a SN~2011fe spectrum with a
similar phase (to within \about 1~day).  Specifically, the $-9.1$,
$-4.7$, $-0.4$, 3.0, and 8.8-day SN~2011by spectra are matched to
$-10.0$, $-3.0$, 0.0, 3.2, and 9.1-day SN~2011fe spectra,
respectively.  With a small phase difference between spectra, spectral
feature differences should be minimal, especially for such similar
SNe.  After correcting for Milky Way reddening and their distances,
each SN~2011by spectrum is dereddened (to account for potential
host-galaxy reddening) and scaled to match its corresponding
Milky-Way-dereddened SN~2011fe spectrum.  Since the phases are not
perfectly matched, we do not expect the flux scaling, which
corresponds to an achromatic offset, to be correct; however, the
colour evolution between these epochs should generally be small,
allowing for an accurate measurement of the reddening.  This method is
graphically outlined in Figure~\ref{f:spec}.

Using a \citet{Fitzpatrick99} reddening law with $R_{V} = 3.1$, we
find that the reddening of SN~2011by (relative to SN~2011fe) ranges
from $E(B-V) = -0.11$~mag (i.e., SN~2011fe is redder) to 0.06~mag,
with 3/5 of the epochs having $E(B-V) = 0.023$ to 0.027~mag.  Only the
first epoch has ``negative reddening,'' which we consider to be the
result of a combination of a quickly changing spectral energy
distribution (SED) at these early times and a phase difference of
0.9~day.  Ignoring the first epoch, the remaining epochs have an
average reddening of $E(B-V) = 0.039 \pm 0.006$~mag, which we consider
to be our best estimate of the host-galaxy reddening of SN~2011by
relative to that of SN~2011fe.  Fixing the reddening to this value,
the best-fitting coherent absolute magnitude offset is $0.303 \pm
0.006$~mag.  We note that the uncertainty on this number neglects the
distance uncertainty and reddening uncertainty (as the reddening was
fixed).

Assuming no host-galaxy dust reddening for SN~2011fe and $E(B-V) =
0.039 \pm 0.006$~mag for SN~2011by, the peak $V$-band absolute
magnitude for each SN is $-19.18 \pm 0.05$~mag and $-18.85 \pm
0.07$~mag, respectively.  The majority of the uncertainties in these
values is related to the distances, as determined from the Cepheid
measurements.  Since the distance estimates are obtained through the
same method, several sources of uncertainty do not apply to a relative
distance measurement.  Taking our best estimates for different terms,
we can determine the relative peak $V$-band absolute magnitude and
uncertainty,
\begin{align}
  \Delta M_{V} &= \Delta m - \Delta E(B-V) R_{V} - \Delta \mu, \\
  \sigma_{\Delta M_{V}} &= \left ( \sigma_{\Delta m}^{2} + \left ( 3.1
      \sigma_{E(B-V)} \right )^{2} + \sigma_{\Delta \mu}^{2} \right )^{1/2},
\end{align}
with parameters and uncertainties
\begin{align}
  (\Delta m, \Delta E(B-V), R_{V}, \Delta \mu) &= (2.93, 0.044, 3.1, 2.449),\\
  (\sigma_{\Delta m}, \sigma_{E(B-V)}, \sigma_{\Delta \mu}) & = ( 0.007, 0.006, 0.062),
\end{align}
where $\Delta E(B-V)$ is the total difference in reddening, including
the Milky Way contribution. We therefore find
\begin{equation}
  \Delta M_{V} = 0.335 \pm 0.063 {\rm ~mag},
\end{equation}
with SN~2011fe being more luminous.  The difference is consistent
with, but smaller than, that found by \citetalias{Foley13:met}: $0.60
\pm 0.36$~mag.  It is also consistent with the \citet{Riess16}
difference (0.286~mag), which only used optical light-curve data.  If
one also makes a 0.06~mag host-mass correction \citep[e.g.,][see
Section~\ref{s:disc}]{Kelly10, Lampeitl10:host, Sullivan10}, there is
no difference between our measurement and that of \citet{Riess16}.
With the updated relative distances, we find that SNe~2011by and
2011fe had peak $V$-band absolute magnitudes which were significantly
different, with a significance of 4.9$\sigma$.  Furthermore, we find a
coherent, reddening-free (Wesenheit) offset of $0.314 \pm 0.062$~mag,
corresponding to a significance of 5.1$\sigma$.

\begin{figure*}
\begin{center}
\includegraphics[angle=0,width=6.4in]{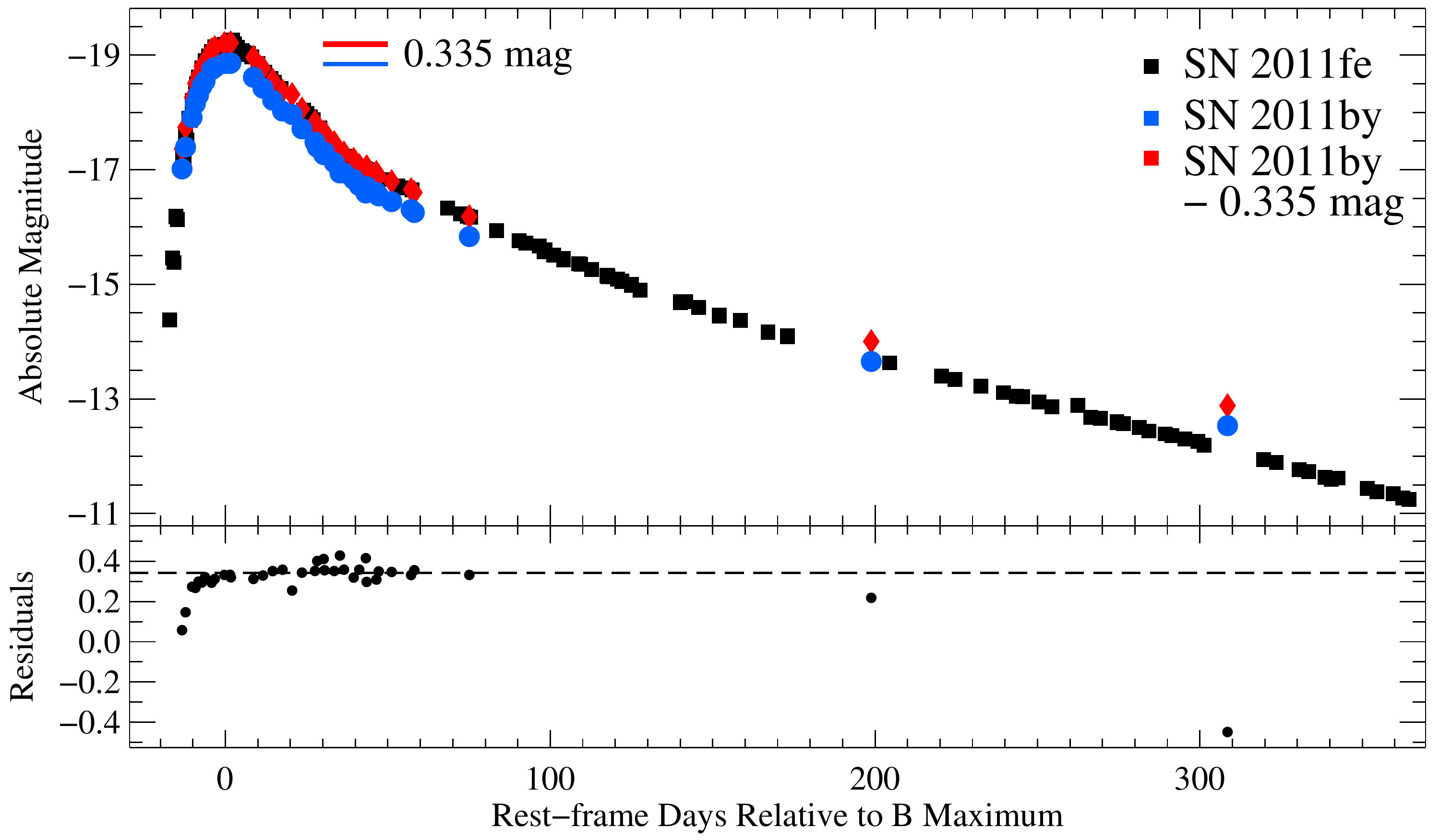}
\caption{{\it Top}: Absolute $V$-band light curves of SNe~2011by (blue
  circles) and 2011fe (black squares) as presented by
  \citetalias{Graham15}, but adjusted to account for the new distances
  to each SN and the measured host-galaxy extinction for SN~2011by.
  The SN~2011fe light curve is a combination of data from
  \citet{Richmond12}, \citetalias{Graham15}, and \citet{Zhang16:11fe}.
  The SN~2011by data are from \citet{Silverman13} and
  \citetalias{Graham15}.  Also plotted is the SN~2011by light curve
  shifted by 0.335~mag, corresponding to the difference in peak
  brightness between the two SNe. {\it Bottom}: Difference between the
  absolute $V$-band light curves for SN~2011by and SN~2011fe, where
  the SN~2011fe light curve was interpolated to match the epochs of
  the SN~2011by observations.  The dashed line is at
  0.335~mag.}\label{f:absv}
\end{center}
\end{figure*}

We present updated absolute $V$-band light curves in
Figure~\ref{f:absv} to illustrate these differences.  This figure uses
an expanded set of light-curve data from a variety of sources (see
\citetalias{Graham15}, and references therein) to show the full
evolution of SNe~2011by and 2011fe for \about 1~yr after explosion.

The grey magnitude offset corresponds to a difference in peak
bolometric luminosity.  SNe~2011by and 2011fe have practically
indistinguishable maximum-light (as well as essentially all other
epochs) optical spectra \citepalias{Foley13:met, Graham15},
light-curve shapes \citepalias{Foley13:met}, and colour curves
\citepalias{Graham15}.  Therefore, their bolometric corrections must
be similar.  While their UV spectra are different, this will have a
minor effect on the bolometric luminosity (at most a few percent); it
is also in the direction of SN~2011fe having a higher bolometric
luminosity than of SN~2011by, which would make their luminosities even
more discrepant.  While it is possible that their mid-IR and far-IR
emission is significantly different, this is unlikely given all other
evidence, and should have a small overall effect on the bolometric
luminosity.

To determine the peak bolometric luminosity, we use the peak spectra
for the two SNe \citepalias{Foley13:met}, which cover the rest-frame
wavelengths \about 1600 -- 10,200~\AA\ for both SNe.  We then extend
the spectra to the NIR by extrapolating an 18,000~K blackbody whose
flux is matched to the optical data.  Since we mostly care about the
relative luminosities, the exact NIR SED is not especially important.
Integrating the spectra and using the distance measurements for each
SN, we find peak bolometric luminosities of $(9.3 \pm 0.7) \times
10^{42}$ and $(12.9 \pm 0.7) \times 10^{42}$~erg~s$^{-1}$ for
SNe~2011by and 2011fe, respectively.

To determine the $^{56}$Ni mass, we use the common formula,
\begin{equation}
  M_{^{56}{\rm Ni}} = \frac{L_{\rm bol,~max}}{\alpha \dot{S}(t_{r, {\rm ~bol}})}, \label{e:mni}
\end{equation}
where $\alpha$ is a unitless parameter of order unity that describes
the diffusion of radiation through the ejecta, $\dot{S}$ is the
instantaneous rate of energy injection from radioactive decay, and
$t_{r, {\rm ~bol}}$ is the bolometric rise time \citep[see
e.g.,][]{Arnett82, Jeffery99}.  For this analysis, we use $\alpha = 1
\pm 0.2$, which is a common choice \citep[e.g.,][]{Stritzinger06:ni}.
Others have chosen a slightly higher value for $\alpha$ \citep[$\alpha
= 1.2 \pm 0.2$;][]{Scalzo10, Pereira13}; however, the exact choice
does not affect our main result, which examines the ratio of $^{56}$Ni
masses.

For this analysis, we use a bolometric rise time of $t_{r, {\rm ~bol}}
= 16.58 \pm 0.14$~days for SN~2011fe \citep{Pereira13}, and assume
that SN~2011by has the same rise time.  This assumption is reasonable
given the similar light-curve shapes and colour curves for the two
SNe.  We note that there is some evidence of SN~2011by having a
slightly slower $V$-band rise (Figure~\ref{f:absv}); however,
\citetalias{Graham15} found that the SN~2011fe rise time was $0.6 \pm
0.4$~days longer than that of SN~2011by (opposite of what one might
immediately assume from the differences in the $V$ band).  Changing
the rise time by a day in either direction only affects the SN~2011by
$^{56}$Ni mass by 0.02~M$_{\sun}$.  Using Equation~\ref{e:mni}, we
find $^{56}$Ni masses of $0.43 \pm 0.09$ and $0.59 \pm
0.12$~M$_{\sun}$ for SNe~2011by and 2011fe, respectively, where the
largest component of the uncertainty is from $\alpha$.

For the above calculation, we propagated the uncertainty from several
different sources (rise time, $\alpha$, etc) that are likely the same
for both SNe.  In the scenario where these values are the same for
both SNe, but the exact value is uncertain, we can measure the ratio
of the $^{56}$Ni masses with an uncertainty which does not include the
uncertainties of these different values.  Doing this, we find a
$^{56}$Ni ratio of $M_{\rm 11fe}(^{56}{\rm Ni}) / M_{\rm
  11by}(^{56}{\rm Ni}) = 1.38 \pm 0.09$.  This value is similar to
that found by \citetalias{Foley13:met} (1.7), but slightly lower,
which is unsurprising given the similar, but slightly different
luminosities for the two analyses.  With the assumptions listed above,
the two SNe have significantly different $^{56}$Ni masses, with
4.2$\sigma$ significance.

\begin{deluxetable}{lll}
\tabletypesize{\footnotesize}
\tablewidth{0pt}
\tablecaption{Supernova and Host-Galaxy Properties\label{t:data}}
\tablehead{
\colhead{Parameter} & \colhead{SN~2011by} & \colhead{SN~2011fe}}

\startdata

Galaxy                                    & NGC~3972               & M101\\
$\mu$ (mag)                               & 31.594 (0.071)         & 29.135 (0.047)\\
$V_{\rm peak}$ (mag)                      & 12.91 (0.01)$^{\rm F}$ & 9.98 (0.02)$^{\rm P}$\\
$\Delta m_{15}(B)$ (mag)                  & 1.14 (0.03)$^{\rm S}$  & 1.10 (0.04)$^{\rm P}$\\
$E(B-V)_{\rm MW}$ (mag)                   & 0.013                  & 0.008\\
$E(B-V)_{\rm host}$ (mag)                 & 0.039 (0.006)          & 0\\
$M_{V, {\rm ~peak}}$ (mag)                & $-18.85$ (0.07)        & $-19.18$ (0.05)\\
Peak Bol.\ Lum.\ ($10^{42}$~erg~s$^{-1}$) & 9.3 (0.7)              & 12.9 (0.7)\\
$^{56}$Ni Mass (M$_{\sun}$)               & 0.43 (0.09)            & 0.59 (0.12)\\
$M_{V, \rm{ ~peak}}$ Offset (mag)         & 0.33 (0.07)            & 0\\
$^{56}$Ni Mass Ratio (Rel.\ to 11by)      & 1                      & 1.38 (0.09)\\
\citetalias{Phillips99} $M_{V, \rm{ ~peak}}$ Offset (mag) & 0.028 (0.022) & 0\\
SALT $\mu$ (mag)                          & 31.96 (0.04)           & 29.16 (0.06)\\
SALT $M_{B}$ Offset (mag)                 & 0.16 (0.08)            & 0\\
Cepheid--SN $\mu$ Offset (mag)            & $-0.14$ (0.08)         & $+0.15$ (0.08)\\

\enddata

\tablecomments{F = \citetalias{Foley13:met}; P = \citet{Pereira13}; S = \citet{Silverman13}.}

\vspace{-0.7cm}

\end{deluxetable}

\subsection{Light-curve Distance Estimates}

In the previous section, we determined that SNe~2011by and 2011fe had
significantly different luminosities and $^{56}$Ni masses.  However,
these SNe do not fall outside the range of all SNe~Ia.  For measuring
distances with SNe~Ia, differences in light-curve shape and colour
make SNe with different intrinsic luminosities have similar {\it
  corrected luminosities}.  SNe~2011by and 2011fe have very similar
observational properties, and thus it would be unlikely for these twin
SNe to have the significantly different corrected luminosities.

None the less, we go through this exercise below.  Importantly, we
must consider the intrinsic scatter for a given distance-fitting
algorithm to determine the significance of any difference.  The
intrinsic scatter indicates how much diversity in distances there is
after correcting for photometric parameters and can be the result of
additional physical conditions such as metallicity.  If SNe~2011by and
2011fe are different by less than the intrinsic scatter, they would
not represent outliers for measuring cosmological parameters.
However, whatever causes the differences in their peak luminosity
would likely contribute to the overall intrinsic scatter measured for
large samples of objects.

The simplest approach to determine the expected difference in $M_{V,
  {\rm ~peak}}$ is to use the decline-rate parameter, $\Delta
m_{15}(B)$.  \citet{Phillips99} determined empirical equations for
this relation that are independent of colour.  For the $V$ band, they
find
\begin{equation}
  \Delta M_{V, {\rm ~peak}} = 0.672 (\Delta m_{15}(B) - 1.1) + 0.633 (\Delta m_{15}(B) - 1.1)^{2},
\end{equation}
where the difference is explicitly relative to a SN with $\Delta
m_{15}(B) = 1.1$~mag.  Since SN~2011fe conveniently has exactly this
value, the equation can be used to determine the expected difference
between SNe~2011by and 2011fe, yielding $0.028 \pm 0.022$~mag, with
SN~2011fe expected to be slightly (insignificantly) more luminous.  As
expected, based on the light-curve shape alone, the two SNe are
expected to have nearly identical peak luminosities.

One can now test how different the relative distances derived from the
light-curve shape and apparent brightness of SNe~2011by and 2011fe are
from those directly measured from Cepheids.  The difference between
these two methods reduces to a difference in peak absolute magnitudes,
as determined by both methods.  This difference is $0.44 \pm
0.11$~mag, including an intrinsic scatter term of 0.09~mag (as
determined by \citealt{Phillips99}) for the uncertainty.  This
difference is 3.8$\sigma$ significant.

We also used the SALT2 algorithm \citep{Guy07} to determine the
distances and peak absolute magnitudes of the two SNe.  SALT2
simultaneously fits multiple light curves, which should include subtle
differences in colour and light-curve shape in any distance estimate.

The difference between SALT2-estimated $B$-band absolute magnitudes is
$0.16 \pm 0.08$~mag, consistent with that found with the
\citet{Phillips99} method, and consistent with zero offset in absolute
magnitude.  Since the SALT2 method corrects for colour differences,
which could be caused by reddening, this value should be compared to
the difference in $A_{B}$ estimated from a direct comparison of the SN
light curves,
\begin{align}
  A_{B} &= R_{B} \times E(B-V) = 4.1 \times (0.039 \pm 0.006) {\rm ~mag}\\ \notag
    &= 0.160 \pm 0.025 {\rm ~mag}.
\end{align}
Therefore, SALT2 is measuring the slight colour difference between the
two SNe, but is incapable of measuring the grey offset between them.
In other words, SALT2 expects SNe~2011by and 2011fe to have the same
intrinsic luminosities.  As a result, the difference in Hubble
residuals (HRs) for the two SNe is approximately the difference
between the reddening-uncorrected $\Delta M_{V}$ and the SALT2
difference: $\Delta {\rm HR} = 0.33$~mag.  This is similar to what was
seen by \citet{Riess16}, finding $\Delta {\rm HR} = 0.29$~mag using
SALT2.

\citet{Riess16} found that SNe~2011by and 2011fe have HRs of $-0.14$
and 0.15~mag, respectively, making neither an outlier relative to the
scatter.  On the other hand, it is perhaps a coincidence that
SN~2011by is faint relative to the mean of the sample and SN~2011fe is
bright relative to the mean of the sample.  If either SN~2011by
(SN~2011fe) were replaced with a true copy of SN~2011fe (SN~2011by),
the mean absolute magnitude of the 19 Cepheid-SN calibrators would
shift by $-0.016$~mag (+0.019~mag).  This shifts the value of H$_{0}$
by 0.72\% to 0.86\%, changing the best-estimate \citet{Riess16} value
of H$_{0}$ of 73.24~km\,s$^{-1}$\,Mpc$^{-1}$ to 72.71 and
73.87~km\,s$^{-1}$\,Mpc$^{-1}$, respectively.  This would be a small
change to the value of H$_{0}$; however, the relatively small size of
the Cepheid-SN calibrator sample, where individual objects can
influence the measured value of H$_{0}$, will clearly be a limiting
factor in the very near future.


\section{Discussion and Conclusions}\label{s:disc}

We find that SNe~2011by and 2011fe, despite having nearly identical
optical spectral sequences \citepalias{Graham15} and light-curve
shapes/colours \citepalias{Foley13:met, Graham15}, have different peak
luminosities.  This result is similar to that of
\citetalias{Foley13:met} --- but with our new Cepheid distance to
NGC~3972, this result is highly statistically significant.  The
observations suggest that SN~2011fe generated 38\% more $^{56}$Ni than
SN~2011by.  Furthermore, the differences in luminosity are
significantly larger than the intrinsic luminosity scatter for SNe~Ia.

The one substantial luminosity-independent observational difference
between SNe~2011by and 2011fe is their near-peak UV flux difference
\citepalias{Foley13:met, Graham15}.  SN~2011fe had relatively more UV
flux and a higher peak bolometric luminosity than SN~2011by, both of
which were predictions for a SN with relatively low progenitor
metallicity \citep{Timmes03, Mazzali06}.  Our current observations are
further indication that the progenitors of SNe~2011by and 2011fe had
super-solar and sub-solar metallicity, respectively
(\citetalias{Foley13:met}; \citealt{Mazzali14};
\citetalias{Graham15}).

If SNe~Ia were perfectly described by their optical light-curve shape
and colour, then SNe~2011by and 2011fe should have the same peak
luminosity.  The difference in luminosities indicates that additional
unaccounted physics must increase the scatter of SN~Ia measurements
beyond errors in analysis.  This comparison indicates that progenitor
metallicity is likely a large component of the ``intrinsic scatter.''
However, current distance estimators do not consider SNe~2011by and
2011fe significantly different (in part because of their internal
model uncertainties and in part because of the intrinsic scatter), and
thus neither would necessarily be excluded from a cosmological
analysis.  Their inclusion as SN-Cepheid calibrators has a minimal
impact on the value of H$_{0}$, but will be limiting in the near
future if the size of the calibrator sample is not increased
significantly.

Recent cosmological analyses have found a significant offset in the
average Hubble residual between SNe~Ia in low-mass and high-mass
galaxies \citep{Kelly10, Lampeitl10:host, Sullivan10}.  Although the
exact functional form of any relation is not known
\citep[e.g.,][]{Childress13}, a correction is often applied as a step
function in host-galaxy stellar masses with the step at
$10^{10}$~M$_{\sun}$.  While this effect has not been significant in
all analyses \citep[e.g.,][]{Scolnic18:ps1}, a typical value for the
difference is \about 0.06~mag, with SNe~Ia in low-mass galaxies being
fainter.  Our results are broadly consistent with this trend, where
NGC~3972 and M101 have stellar masses below and above
$10^{10}$~M$_{\sun}$, respectively.  It is therefore possible that the
mass step is driven primarily by metallicity differences in the
progenitor systems which correlate with the host galaxy's total
stellar mass at the time of explosion.

A recent analysis of ``twin'' SNe~Ia suggested a low intrinsic
dispersion of only 0.08~mag \citep{Fakhouri15}.  SNe~2011by and 2011fe
are, by all accounts, better twins than those presented by
\citet{Fakhouri15}, yet they have a peak absolute magnitude difference
that is \about 4 times larger than the scatter for their sample.
While choosing SNe~Ia with similar spectral properties should not
typically {\it increase} scatter, the incredibly well-observed and
precisely measured SNe~2011by and 2011fe provide a cautionary tale for
using this method for improving distance estimates.

Future analyses must evaluate how this result will affect their
conclusions.  Obviously, there is additional information in the UV
SED, and observing the rest-frame UV could potentially improve
SN~Ia distance estimates.  However, if the average progenitor
metallicity is changing with redshift \citep[e.g.,][]{Childress14},
there may be a systematic bias to SN~Ia distances with redshift.  As
there is some indication of changing UV properties with redshift
\citep{Foley12:sdss, Maguire12, Milne15}, a thorough analysis of this
effect should be performed.  In particular, additional UV spectra of
low-redshift SNe~Ia will be critical for understanding how progenitor
metallicity affects SN~Ia distance estimates.

\section*{Acknowledgements}

{\it Facility:} {\it HST} (ACS, STIS, WFC3)

\bigskip

Based on observations made with the NASA/ESA {\it Hubble Space
  Telescope}, obtained at the Space Telescope Science Institute
(STScI), which is operated by the Association of Universities for
Research in Astronomy, Inc., under National Aeronautics and Space
Administration (NASA) contract NAS 5--26555. These observations are
associated with Programmes GO--12298 and GO--13647.  Support for
GO--13647 was provided by NASA through a grant from STscI.  This
manuscript is based upon work supported by NASA under Contract No.\
NNG16PJ34C issued through the {\it WFIRST} Science Investigation Teams
Programme.

The UCSC team is supported in part by NASA grant NNG17PX03C, NSF
grant AST-1518052, the Gordon \& Betty Moore Foundation, the
Heising-Simons Foundation, and by fellowships from the Alfred P.\
Sloan Foundation and the David and Lucile Packard Foundation to R.J.F.

P.J.B./the {\it Swift} Optical/Ultraviolet Supernova Archive (SOUSA)
and P.A.M.\ are supported by NASA's Astrophysics Data Analysis Program
through grants NNX13AF35G and NNX17AF15G, respectively. A.V.F.\ is
grateful for assistance from the Christopher R. Redlich Fund, the
TABASGO Foundation, and the Miller Institute for Basic Research in
Science (U.C. Berkeley).  KAIT and its ongoing operation were made
possible by donations from Sun Microsystems, Inc., the Hewlett-Packard
Company, AutoScope Corporation, Lick Observatory, the NSF, the
University of California, the Sylvia \& Jim Katzman Foundation, and
the TABASGO Foundation.


\bibliographystyle{mnras}
\bibliography{../astro_refs}


\end{document}